\documentclass[a4paper,12pt]{article}
\pdfoutput=1

\usepackage[tbtags]{amsmath}
\usepackage{amsfonts,amssymb,amsthm}
\usepackage{eucal}
\usepackage{setspace}
\usepackage[round]{natbib}
\usepackage{graphicx}
\usepackage{verbatim}
\usepackage{url}
\usepackage{lineno} 
\usepackage{enumerate}
\usepackage{bm}
\usepackage{soul}
\usepackage{authblk}
\usepackage{dsfont}
\newcommand{\R}{\mathds{R}}

\usepackage{xspace}
\newcommand{\Matern}{Mat\'{e}rn\xspace}

\newcommand{\given}{\mathbin{\vert}\nolinebreak}
\newcommand{\ldef}{\mathrel{:=}\nolinebreak}
\newcommand{\trans}{^{\scriptscriptstyle T}}
\newcommand{\bzero}{\boldsymbol{0}}
\newcommand{\rd}{\textrm{d}}
\newcommand{\code}[1]{\texttt{#1}}
\newcommand{\unit}[1]{\ensuremath{\,\mathrm{#1}}}
\newcommand{\half}{\frac{1}{2}}

\newcommand{\curlyA}{\mathcal{A}} 
\newcommand{\curlyB}{\mathcal{B}} 
\newcommand{\Xtilde}{\widetilde{X}}
\newcommand{\Ytilde}{\widetilde{Y}}
\newcommand{\ytilde}{\tilde{y}}
\newcommand{\Ztilde}{\widetilde{Z}}
\newcommand{\ztilde}{\tilde{z}}
\DeclareMathOperator{\pr}{p}
\DeclareMathOperator{\bigO}{O}
\DeclareMathOperator{\E}{\mathds{E}}
\DeclareMathOperator{\Var}{\mathds{V}}
\DeclareMathOperator{\CV}{CV}

\title{Bayesian model-data synthesis with an application to global Glacio-Isostatic Adjustment}

\author[1,2]{Zhe Sha}
\author[2]{Jonathan Rougier}
\author[1]{Maike Schumacher}
\author[1]{Jonathan Bamber}
\affil[1]{School of Geographical Sciences, University of Bristol}
\affil[2]{School of Mathematics, University of Bristol}
\date{}
\begin{document}

\maketitle


\begin{abstract}

We introduce a framework for updating large scale geospatial processes using a model-data synthesis method based on Bayesian hierarchical modelling. Two major challenges come from updating large-scale Gaussian process and modelling non-stationarity. To address the first, we adopt the SPDE approach that uses a sparse Gaussian Markov random fields (GMRF) approximation to reduce the computational cost and implement the Bayesian inference by using the INLA method. For non-stationary global processes, we propose two general models that accommodate commonly-seen geospatial problems. Finally, we show an example of updating an estimate of global glacial isostatic adjustment (GIA) using GPS measurements.

\end{abstract}


\section{Introduction}


This paper presents a model-data synthesis method based on Bayesian
hierarchical modelling \citep[BHM, see, e.g.,][]{banerjee04,cressie11}.
The main feature of our approach is to use observations to adjust
a model-based solution (`simulation'), by modelling explicitly the
discrepancy between the simulation and the true process.  This approach
follows a long tradition in the field of computer experiments, in
which such a discrepancy is a key part of the statistical
model linking simulations, reality, and observations \citep[see,
e.g.,][]{koh01,cgrs01,gr04,gr09reify}.  In contrast, a popular
alternative is an approach  in which the true process is modelled explicitly, and
the simulation and observations are both treated as measurements;
see \citet{rougier13mme} and the references therein, for the use of
this approach in climate science.  There are also hybrid approaches, such as
\citet{zammit14}.

In spatial
applications there are several advantages to modelling the discrepancy
explicitly, as opposed to modelling the underlying process.
First, it is more defensible to treat the discrepancy in a parsimonious way as an
expectation-zero isotropic stochastic process, with---one hopes---a
relatively short correlation length.  This presumes that the
simulation gets most of the large-scale `non-stationary' features right,
in which
case the discrepancy concerns itself mainly with smaller-scale local
features.  Second, the discrepancy is often the primary concern of
researchers, who are interested in where their simulation performs
poorly, and how large these regions are.  Likewise, in comparing
simulations, discrepancy maps can shed light on how different
representations of process components cause different types of error.

The outline of the paper is as follows.  Section~\ref{sec:model}
describes our general approach to synthesizing a simulated field
and observations, and the challenges of performing this operation at scale.
We also describe the stochastic partial differential equation approach to
exploiting the sparsity of Gaussian Markov random fields.
Section~\ref{sec:nonstat} describes two different approaches to
modelling non-stationary prior beliefs, and some useful variants.
Section~\ref{sec:GIA} describes an application to global isostatic
adjustment, based on a one-degree resolution simulation and 
2500 GPS stations.  Section~\ref{sec:conc} is a brief conclusion.

\section{Model and methods}
\label{sec:model}

\subsection{Outline of our approach}

The outline of our approach is easily stated, although, for
computation at scale, the devil is in the details.  Without loss
of generality, the process domain is taken to be the surface of the
Earth, denoted $\mathds{S}^2$.  The simulation and the true process are
both functions $\mathds{S}^2 \to \R$; we denote the simulation as $m$,
with a small letter because it is known, and the process as $X$,
with a large letter because it is unknown.  For clarity, we will refer to $X$ throughout as the `latent process'.
The stochastic process representing uncertainty about 
$X$ is induced through the specification of a stochastic process for
the discrepancy $\Xtilde \ldef X - m$.  The observations $y \ldef (y_1,
\dots, y_n)\trans$ are measurements on known linear maps of $X$, 
possibly made with error.  The simulation/observations synthesis
involves conditioning $\Xtilde$ on $y$.  After conditioning, the updated discrepancy process ${\Xtilde \given y}$ can be summarised in terms of point values, areal averages,
or differences, as required, each quantified in terms of a posterior
expectation, with uncertainty summarized in terms of a  posterior
standard deviation.  If necessary, the latent process $X$ can be reconstructed by adding $m$ back to ${\Xtilde \given y}$.

In this approach, the most demanding aspect is the specification of
the stochastic process for the discrepancy $\Xtilde$.  In contrast,
the relationship between the latent process $X$ and the observations
$y$ is typically well-understood, and the observation error reasonably
well-quantified.  We tackle the challenge of specifying a stochastic
process for $\Xtilde$ in two steps.  First, we treat $\Xtilde$ as
expectation-zero, isotropic, and Gaussian.  These strong modelling
assumptions are very common in spatial statistics, but vary in their
defensibility.  In our approach, we defend them on the basis that we
are modelling the discrepancy between simulation and latent process, for
which we can expect the stochastic structure to be far simpler than
the approach where we are modelling the latent process explicitly.
Essentially, we are placing our faith in the simulation $m$, to
have got the large-scale features of the latent process about right.
Second, we do not commit to a specific choice of expectation-zero,
isotropic, and Gaussian process, but allow its two parameters, the
marginal variance~$\sigma^2$ and the range parameter~$\rho$ to
vary within some parametric family.  We write $\theta \ldef (\sigma^2,
\rho)$, and term these the `hyperparameters'.

Now we can summarize our statistical model in the standard hierarchical form.  We
introduce known linear maps $\curlyA \ldef (\curlyA_1, \dots,
\curlyA_n)\trans$, for which $Y_i = \curlyA_i X + E_i$, where $E_i$
is measurement error, and $E$ is treated as probabilistically independent of $X$.
Then the observations are modelled as
\[
  Y = \curlyA X + E ,
\]
where $E \sim \text{N}_n(\bzero, Q_Y^{-1})$, and $Q_Y$ is a known precision matrix, typically sparse and often diagonal.  The latent process is modelled as
\[
  X = m + \Xtilde
\]
where the stochastic process $\Xtilde$ is parametrized by the hyperparameters $\theta$.  After rearranging, the final hierarchical
form of the model is
\begin{subequations}\label{eq:hier1}
\begin{align}
  \Ytilde \given \Xtilde, \theta & \sim \text{N}_n(\curlyA \Xtilde, Q_Y^{-1}) \\
  \Xtilde \given \theta & \sim \text{GP}(\theta) \\
  \theta & \sim \pi
\end{align}
\end{subequations}
where $\Ytilde \ldef Y - \curlyA m$, `GP' denotes an expectation-zero
isotropic Gaussian process on $\mathds{S}^2$, and $\pi$ is a prior
distribution for the hyperparameters. In this formulation, the simulation $m$ is used to adjusted the observation $Y$ prior to update the discrepancy process $\Xtilde$. If necessary, the hyperparameters can be extended to include parameters for $\curlyA$ and $Q_Y$.

\begin{figure}[htbp]
 \centering
\includegraphics[width=0.9\textwidth]{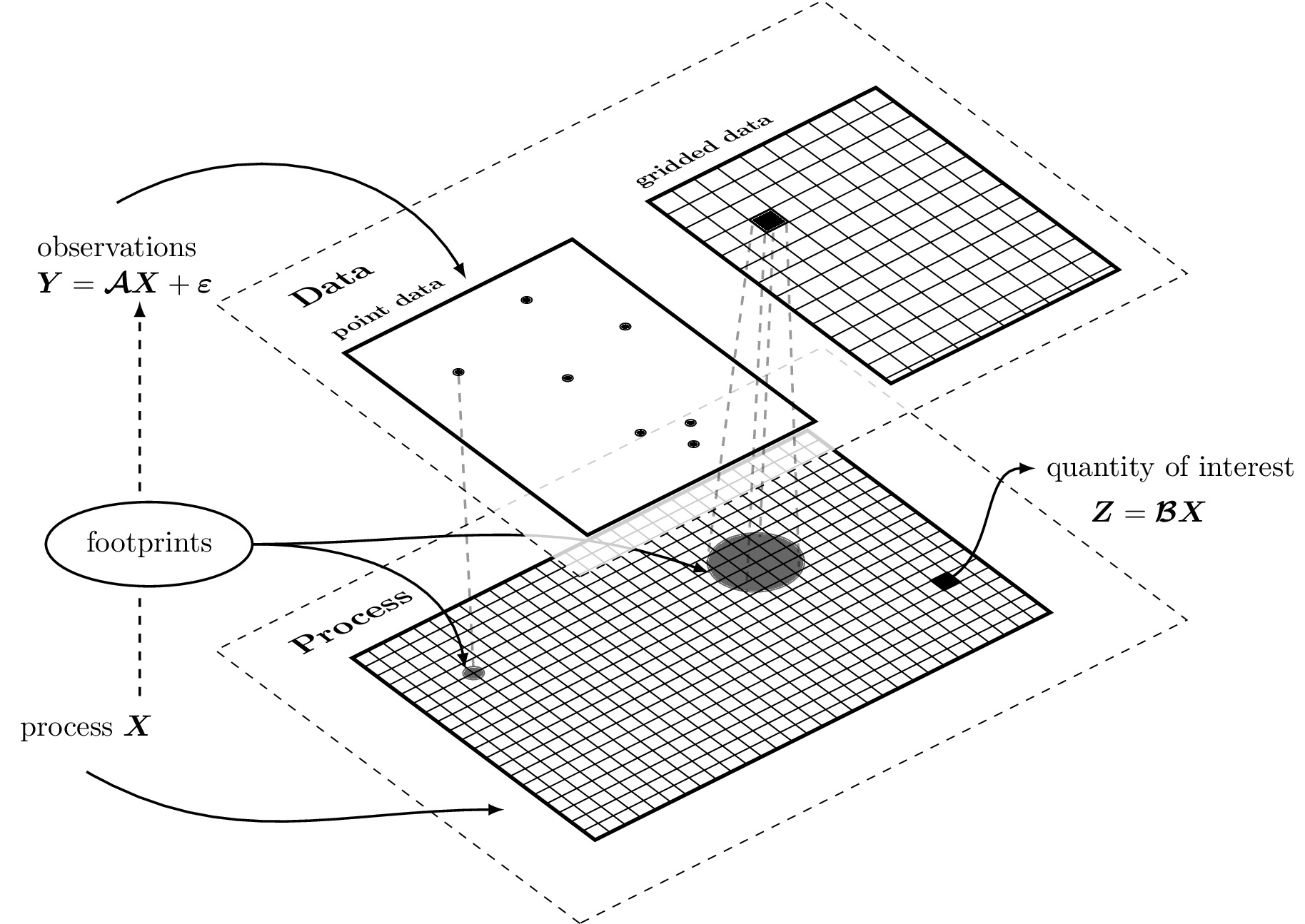}
\caption{Framework of the Bayesian model-data synthesis method.}
\label{fig:framework}
 \end{figure}

To summarize the updated discrepancy process we introduce additional
known linear maps $\curlyB \ldef (\curlyB_1, \dots,
\curlyB_m)\trans$, for which $\Ztilde \ldef \curlyB \Xtilde$ are
the quantities of interest.  The marginal posterior distribution for
$\Ztilde$ factorizes as
\begin{equation}\label{eq:pr*z}
\begin{split}
  \pr^*(\ztilde)
  & = \int \pr^*(\ztilde, \theta) \, \rd \theta \\
  & = \int \pr(\ztilde \given \ytilde \,; \theta) \, \pr(\theta \given \ytilde) \, \rd \theta ,
\end{split}
\end{equation}
where the first term in the integrand has a closed-form expression
which depends on $y$ and on $\curlyA$, $\curlyB$, $Q_Y$, and $\theta$,
while the second term is unlikely to have a closed-form expression.  This type of integration can be approximated using the method of Integrated Nested Laplacian Approximations \cite[INLA, see][]{rue09,martins13}. Figure \ref{fig:framework} shows a graphical illustration of the model outline.

\subsection{Computational issues} \label{sec:GMRF}

We now review the challenge of computing $\pr^*(\ztilde)$, and recent
developments.  Putting aside the integration over $\theta$, the
challenge for
computing \eqref{eq:pr*z} is that the closed-form expression
$\pr(\ztilde \given \ytilde \, ; \theta)$ requires $\bigO(n^3)$ flops to
compute, this being the dominant cost when factorizing the variance
matrix of $\Ytilde$.  Current
desktop technology tends to grind to a halt for $n$ much larger
than a thousand.  This is not nearly enough for some applications,
particularly those where the latent process is defined on the whole
of the surface of the Earth, for which there may be many thousands
of point or areal measurements.  Long computing time is also a major
bottleneck during code development.

There are some simple work-arounds.
Thinning or aggregating the observations down to less than a thousand
is one possibility.  Another is splitting the update into separate regions, each containing less than a thousand observations.  These are valuable
pragmatic approximations, but can be risky when the range parameter of the discrepancy is uncertain, because the value of the range parameter
affects the accuracy of thinning, aggregating, or splitting.  \citet[sec.~1]{lindgren11} discuss other more technical approaches.

Therefore we looks for a different type of approximation, which reduces
the dominant cost to below $\bigO(n^3)$.  Gaussian Markov random fields
(GMRFs) are one answer.  As discussed in \citet{rue05}, GMRFs defined
on a finite-dimensional random vector can exploit sparsity in the
precision matrix.  This sparsity is represented by an undirected graph
where the vertices are the elements of the vector, and the absence of
an edge between two vertices indicates a conditional independence,
usually induced by a neighbourhood structure.  The difficulty of
applying a GMRF approach directly is that $\mathds{S}$ would need to be
discretized into a finite number of pixels.  Having done this,
though, the natural neighbourhood structure would link pixels with
a common boundary (a first-order neighbourhood scheme), for which the computational cost is
$\bigO(n^{3/2})$.  However, it is not straightforward to configure the
precision matrix that results to approximate an isotropic Gaussian
process with variance and range parameters.  It is also somewhat
arbitrary to pixelate the continuous domain $\mathds{S}$, and wasteful
to do so if some regions of $\mathds{S}$ are more interesting and more
highly-observed than others.

The breakthrough came with \citet{lindgren11}; see also \citet{simpson12}, although be warned that the notation in these two papers is at variance.  Peter Whittle had shown that an
isotropic Gaussian process with a \Matern covariance function arose as
a solution to a particular stochastic partial differential equation (SPDE).
\citeauthor{lindgren11}\ were able to use this insight to construct
a finite-dimensional approximation to a \Matern Gaussian process for $\Xtilde$ of the form
\begin{equation}\label{eq:Xtildeapprox}
  \Xtilde(s) \approx \sum_{j=1}^k W_j \, \psi_j(s) ,
\end{equation}
in which the $\psi_j$ are specified according to a triangulation
of $\mathds{S}$, which can be adapted to the needs of the application.
\citeauthor{lindgren11}'s particular choice of $\psi$ induces a GMRF specification
for the vector $W \ldef (W_1, \dots, W_k)^T$, with a sparse precision matrix $Q_W$ with a simple parametrization in terms of $(\sigma^2, \rho)$, where $\rho$ is now
interpreted as the range parameter for a \Matern covariance function. Without this simple specification in terms of the hyperparameters, most of the computational advantages would be lost.
\citet{lindgren11}\ and \citet{simpson12} show that the approximation
error in \eqref{eq:Xtildeapprox} is $\bigO(h)$ where $h$ is a measure
of the triangle size, such as the radius of the largest inscribed
circle, or the longest edge.  Thus the hierarchical model from \eqref{eq:hier1} becomes,
under the approximation in \eqref{eq:Xtildeapprox},
\begin{subequations}\label{eq:hier2}
\begin{align}
  \Ytilde \given W, \theta & \sim \text{N}_n(A W, Q_Y^{-1}) \\\label{eq:hier2b}
  W \given \theta & \sim \text{N}_k\big( \bzero, Q_W^{-1}(\theta) \big) \\ 
  \theta & \sim \pi
\end{align}
\end{subequations}
where $A$ is $(n \times k)$ with $A_{ij} \ldef \curlyA_i \, \psi_j$, and $Q_W(\theta)$ is a
sparse precision matrix with a simple parametrization in terms of $\theta = (\sigma^2, \rho)$.

When combined with sparsity in $Q_W$, sparsity in $A$ and $Q_Y$ (which is often diagonal)
implies sparsity in the precision matrix of ${W \given \{Y ;
\theta\}}$, reducing the cost of conditioning from $\bigO(n^3)$
to something more like $\bigO(n^{3/2})$.  Sparsity in $A$ arises
naturally when the observations have small spatial footprints, because
the $\psi_j$ are localized in the triangulation, and thus also
have small spatial footprints.  Any observation $Y_i$ whose footprint
$\curlyA_i$ does not
overlap $\psi_j$ has $A_{ij} = 0$.  The interplay between the small
footprints in $\curlyA$ and the basis functions $\psi$ is discussed
in more detail in \citet[sec.~3.6]{simpson12}.  The more complex
interplay between $\curlyA$, $\curlyB$, and $\psi$ is discussed in
\citet{zammit18}.

\section{Modelling non-stationarity}\label{sec:nonstat}
In many applications, the latent process is only locally stationary and this may also be true for the discrepancies between the process and the simulations. The nature of the non-stationarity in geophysical processes usually falls in to three categories: (1) the process is defined only on a subset of the domain; (2) the process shows strong regional differences; (3) prior knowledge of the process varies by region. The SPDE approach outlined in Section~\ref{sec:GMRF} can be extended to non-stationary processes, notably those defined by regions with different behaviours.

Before introducing the models for non-stationary processes, we need to
consider the triangulation for the SPDE approximation. Let the domain
be partitioned as regions $\{\Omega_i\}_{i=1}^p$, where we assume that
the boundaries between the regions are made up of linear segments.
If we respect the boundaries by using their segments as triangle edges, then
automatically-generated triangulations can have short edges and sharp
angles, which is inefficient.  On the other hand, if we triangulate the whole domain 
first in an efficient manner, and assign each triangle to a region
according to its centroid, then we distort the boundaries between
regions, and have less control over the resolution in each region.

\begin{figure}[p]
 \centering
\includegraphics[width=0.7\textwidth]{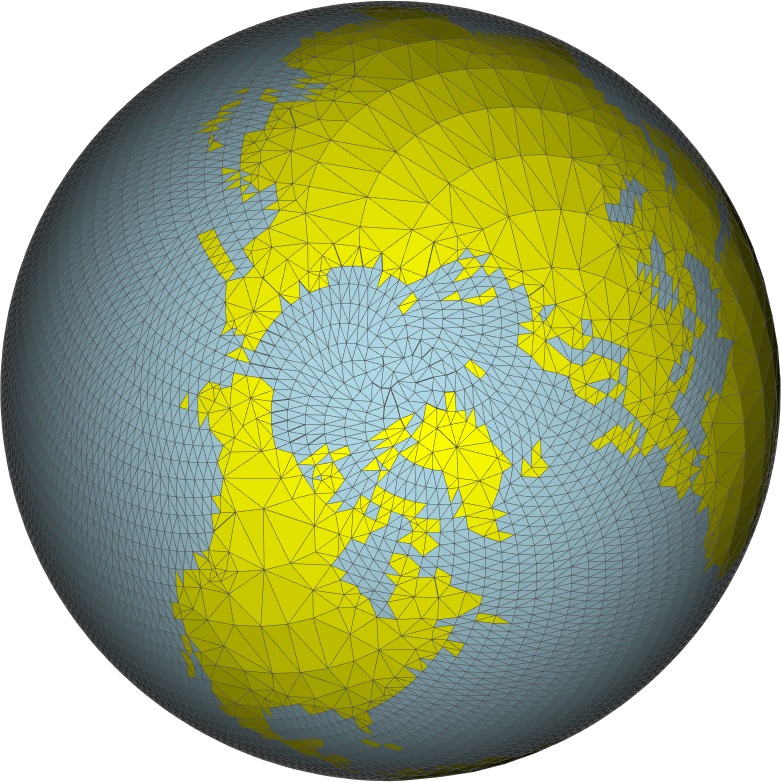}
\caption{Separating triangles over the land and oceans. Here we have used a lower resolution triangulation over the land.}
\label{fig:meshglobe}
 \end{figure}

The approach we adopt below is a compromise, in which we tolerate
a small amount of distortion in order to derive an efficient
triangulation of the whole domain, and variable resolution in the
regions.  We create a vertex set for each region (varying the resolution as appropriate), and
then merge these sets, to cover the entire domain.  Then we
re-triangulate based on the vertices, to control for short edges and
sharp angles.  Finally, we assign each triangle to a region according
to its centroid, and in this way we distort the regions slightly.  If there is too much distortion we can modify the approach, for example by increasing resolution near the boundaries. The \texttt{R} code for triangulation and building the corresponding precision matrices is based on \cite{Haakon2016}. 
Figure \ref{fig:meshglobe} shows an example of this approach to
create a triangulation of the Earth with high resolution over the
oceans and low resolution over land.

\subsection{Subset model}\label{sec:subset}
Some processes are defined only on particular regions. For example, sea level change is only meaningful over the ocean and around the coastal regions. The domain of this process is connected globally by the oceans but separated locally by the lands. For sea level changes at any two points separated by land, their correlation is more likely to depend on the path connecting the two points along the coastline rather than the Euclidean distance across the land. Thus the land introduces non-stationarity.  

In this case, it is natural to model the process only on the subset of interest, and we call this approach the \emph{subset model}. Denote by $\Omega_s \subset \mathds{S}$ the subset of the domain where the process is defined. The GMRF approximation of the process can be built in the same way as the global stationary model using a triangulation over $\Omega_s$ only. Then subset model retains the same form as the BHM in \eqref{eq:hier2}. 

\begin{figure}[htbp]
 \centering
\includegraphics[width=\textwidth]{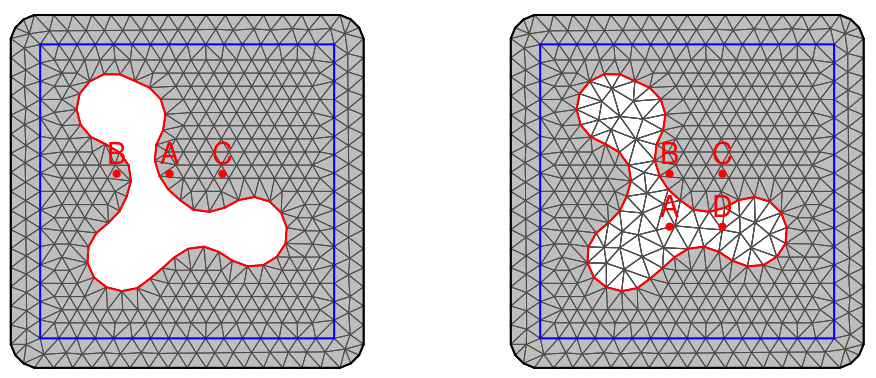}
\caption{An example of GMRF approximation for non-stationary models. 
Left panel (subset model): the Euclidean distances of $AB$ and $AC$ are the same and the two correlations are both $0.19$ in the stationary model. However in the subset model, the correlation is almost zero for $AB$, and $0.27$ for $AC$. 
Right panel (partition model): the shaded region in the middle has $\rho_0 = 1.5$ and the outside has $\rho_s = 1$. The four points $ABCD$ are the vertices of a square. The correlation between $AB$ is about $0.23$, between $AD$ is $0.28$, between $BC$ is $0.14$.}
\label{fig:subpart}
 \end{figure}

The left panel of Figure \ref{fig:subpart} provides an illustration of the subset model on the plane. The domain is a square with side length equal to $5$. Suppose the process has a correlation length equal to $1$, then for an adequate finite element approximation, the maximum edge length of the mesh triangles within the study region should be smaller than 1: we choose the maximum edge length to be $0.5$. The process is defined over the whole square except for the blank region $\Omega_0$ in the middle. Consider the correlation between two pairs of points $AB$ and $AC$ as shown in the plot. The Euclidean distances between them are the same, so in a stationary model they both have correlations around $0.19$. However, in the subset model, the correlation between $AB$ is almost zero ($\approx 5 \times 10^{-7}$), which reflects the fact that $A$ and $B$ are separated by $\Omega_0$; informally, the path in this region is far longer. The correlation between $AC$ is $0.27$. This is slightly higher than the stationary model ($0.19$) because after removing $\Omega_0$, the rest of the triangles get more weights in building up the precision matrix.

\subsection{Partition models}
There are also processes which are well-defined over the whole domain, but which have varying spatial characteristics. A typical example is an atmospheric process that shows very different behaviours across the oceans, coasts and land. We propose \emph{partition models} to capture such spatial heterogeneity. First the domain is partitioned into $p$ regions such as oceans, coastal regions and land. As before, denote this partition by $\{\Omega_i\}_{i = 1}^{p}$. Then there are two approaches. The first is a decomposition of the process according to the region, hence we call it a \emph{process partition model}. The second uses a single process over the domain but varies the hyperparameters according to the region, and therefore we call it a \emph{parameter partition model}.

In a process partition model, the process is decomposed into independent sub-processes defined on each $\Omega_i, i = 1, \dots, p$. If each observation is associated with exactly one $\Omega_i$, then independence in the prior distributions for each $\Omega_i$ implies independence in the posterior distribution as well.

In contrast, the parameter partition model allows correlation between the regions in the prior distribution, and therefore in the posterior distribution as well. The latent process over the domain is modelled as a single Mat{\'e}rn Gaussian process, but the hyperparameters $\theta$ become a function of the location $s$:
\begin{align}
\theta(s; \theta_1, \dots, \theta_p) = \theta_i, \qquad i : s \in \Omega_i.
\end{align}
The parameter partition model takes exactly the same form as the global stationary model in \eqref{eq:hier2} but there are $p$ sets of $\theta_i$ in the covariance function to be estimated. 

The right panel of Figure \ref{fig:subpart} illustrates the spatial correlation in a parameter partition model. The four points $ABCD$ are the vertices of a square. Instead of removing $\Omega_0$ in the middle, we assume the process has a correlation length $\rho_0 = 1.5$ within this region, and $\rho_s = 1.0$ outside it, as before. The resolution of the triangulation within this region can remain the same as the outside region since the correlation length is longer; however we choose to use larger triangles to reduce the computational cost. The correlation between $AD$ is $0.28$ and the correlation between $BC$ is $0.14$. For $AB$ and $CD$ , which cross the boundary,  the correlations are $0.23$ and $0.18$.

\subsection{Constrained partition models}
\label{sec:constrained}

In the above examples, the partition of the spatial domain is determined by physical boundaries. It is also possible to define the regions based on our knowledge and interests. In fact the latter is quite common as human activities and studies are not evenly distributed over the Earth. Therefore we might be more certain about the behaviour of a process in particular regions, and we can use this knowledge to our advantage, to reduce the number of hyperparameters.

We consider one common case in more detail, which will also feature
in the illustration in Section~\ref{sec:GIA}.  Consider a region
where the process is known to high accuracy: for simplicity,
and without loss of generality, suppose the process is known to
be near-zero throughout the region, which  we therefore term the
`zero-region', and denote as $\Omega_0$.  We can use a combination
of hyperparameters and `pseudo-observations' to enforce both the
near-zero value in $\Omega_0$, and also the continuity of the process
across the boundary of $\Omega_0$.  First, we set the correlation
length for $\Omega_0$, so that it is at least half the length
of the longest diagonal.  Then we introduce pseudo-observations
each with value zero and very small error, roughly equally-spaced inside $\Omega_0$.
When combined with the long correlation length for $\Omega_0$,
conditioning on these observations has the effect of constraining the
process inside $\Omega_0$ to be near-zero, and constraining the process
just outside $\Omega_0$ to be close to zero.  This holds regardless of the variance for $\Omega_0$, and therefore a common variance hyperparameter can be used both inside and outside $\Omega_0$.

This approach is more attractive than the alternative of only
modelling the process outside the zero-region, for two reasons.
The first has already been noted: continuity across the boundary of
the zero-region.  The second is that the zero-region is incorporated
within the inference, and does not have to be treated separately.
This simplifies the code and the packaging of results,
and reduces the possibility of error.  The additional cost of more
vertices is slight, because the triangulation in the zero-region can
be low-resolution.

\section{Application to GIA}
\label{sec:GIA}

We apply our approach to update glacial isostatic adjustment (GIA) using global positioning system (GPS) data. GIA represents the very slow vertical movement in the Earth's crust due to the unloading of ice-sheets since the last glacial maximum (about 20 thousand years ago). This movement affects the Earth's geoid and is crucial for interpreting or predicting sea level changes. Over time-intervals as short as decades, GIA change can be treated as a spatially-varying but time-invariant trend, measured in mm/year. 

\cite{guo2012} provide a comprehensive review and comparison of 14 GIA simulations. Typically these simulations are from physical models with assumptions about the Earth's structure and ice-loading history. These simulations often have substantial regional discrepancies from observations due to limitations in the model assumptions. In this application, we use one of the latest GIA simulations from the ICE6G-VM5 model \citep{Peltier}, as our prior expectation $m$ for the GIA process.

GPS observations provide a direct measure of the the vertical rate at point locations in space. Our GPS data are from a global network of 4072 GPS stations for the period 2005-2015 \citep{ms2018}. The data are processed to include only the vertical bedrock movement so that all non-GIA effects and artefacts are removed. Then an annual trend is estimated for each station to represent the observed vertical movement rate. The estimated standard errors of these trends are used for GPS measurement errors. 

Our task is to assimilate the ICE6G simulation and the GPS data into an updated assessment of GIA, with predictive uncertainties.

\subsection{Outline of the calculation}

We want to predict GIA at roughly $100 \,\mbox{km}$ resolution. The simulated GIA solution from the ICE6G model is provided on a standard one-degree longitude-latitude grid. This grid is non-regular in shape because it has a pixel size of about $100 \,\mbox{km}$ at the equator and less than $10 \,\mbox{km}$ at high latitude.

In our approach, the triangulation used in the GMRF representation needs to have approximately the same resolution as the prediction required, and for efficiency the triangles should have similar sizes and shapes. Therefore, we generate an equal area lattice on the sphere using 30,000 Fibonacci points \citep[see, e.g.,][]{Gonzalez2009} as the starting vertices. Euler's formula then implies that this gives about 60,000 triangles with an average spacing of $100 \,\mbox{km}$.

As explained after \eqref{eq:hier1}, we need to adjust the observations by the prior mean field $m$. Thus, the ICE6G gridded values are mapped to the GPS station locations by linear interpolation, and then subtracted from the corresponding GPS values. This gives values for $\Ytilde$ in \eqref{eq:hier1}. Then we use INLA to update the discrepancy $\Xtilde$ on the triangulation by conditioning on $\Ytilde$ and integrating out the hyperparameters $\theta$ according to our prior given below in Section \ref{sec:prior}. The updated discrepancy field is then mapped back to the ICE6G grid, and the GIA process is reconstructed by adding back the simulation $m$. The result is an object similar to a GIA simulation, but representing the posterior expectation of GIA.  Additionally, each pixel in the result is accompanied by a measure of uncertainty, the predicted standard deviation. 

\subsection{Non-stationarity}

GIA is modelled as a single process over the entire Earth, but we are much more certain about its value in some regions than others. For example, in the middle of low latitude oceans, GIA is known to be negligible. However, for various reasons, most simulations from physical models have deviations from zero in these regions, and there is little reliable GPS data available to correct these systematic errors.

Although we could specify these zero-GIA regions `by hand', we prefer a procedure based on the ensemble of GIA simulations already available. To generate the regions, we used the ICE6G solution and $12$ of the $14$ GIA solutions compared in \cite{guo2012}. Five of them are removed as they are out-dated or have obvious flaws. Then the regions are generated by the following procedure. 

\begin{enumerate}[1.]

\item  Calculate the ensemble means and standard errors for each $1
\unit{degree}$ pixel from the eight GIA simulations.

\item Retain the pixels that have values
smaller than $0.3 \unit{mm/year}$, after exploring a few different
thresholds with our experts.

\item Identify and remove dubious pixels by setting threshold values
for the ensemble standard deviation. We remove pixels with standard
deviations larger than $0.4 \unit{mm/year}$.

\item Connect the remaining pixels into polygons and remove polygons
that have area smaller than $200 \unit{km^2}$, which we regard as
too small to be defined as a region.

\end{enumerate}

\begin{figure}[htbp]
 \centering
\includegraphics[width=\textwidth]{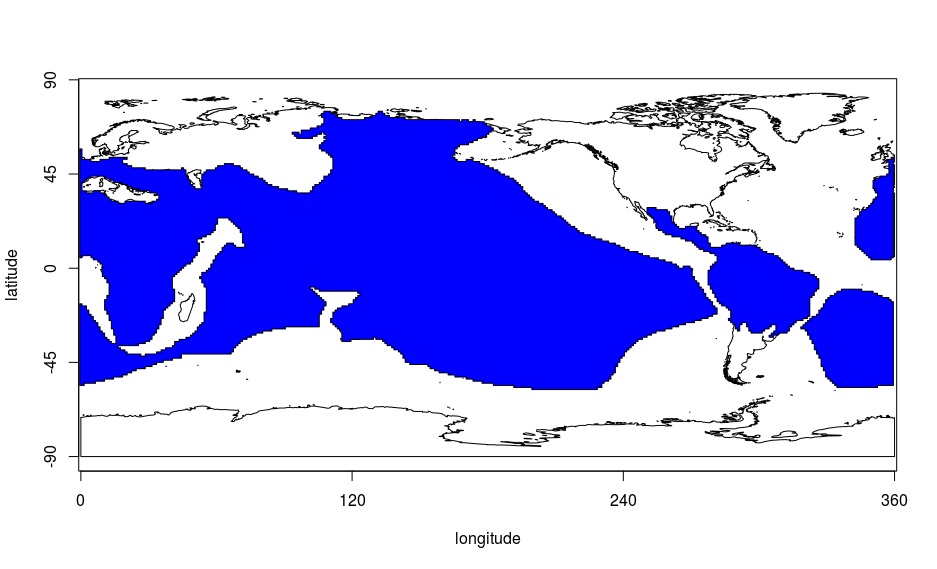}
\caption{The polygons where the GIA is expected to be zero (coloured in blue).}
\label{fig:zeroPoly}
 \end{figure}

Figure \ref{fig:zeroPoly} shows the result of this procedure. We call the union of these polygons the `zero-region' and the complement the `region of interest'. The zero-region contains most of the ocean basins, plus some low-latitude lands, as would be expected.  We also removed the GPS observations inside the zero-region since they are not required for the purpose of this study. The final GPS dataset contains 2515 observations, all in the region of interest.

Given the nature of our \textit{a priori} non-stationarity,
we prefer to use the constrained partition model described in
Section~\ref{sec:constrained}.  For the zero-region, we choose
a correlation length equal to the diameter of the Earth,
which is long enough for a sparse representation of the zero-region and also keeps the precision matrix of the latent process from being singular. We use $50$
pseudo-observations with an error of $0.1 \unit{mm/yr}$, which is
about the same size as the smallest GPS measurement errors, spread
evenly through the zero-region.

By way of contrast, we also present two other approaches: modelling
the process on the entire Earth as \textit{a priori} stationary, and
modelling only the region of interest, as \textit{a priori} stationary.
In the latter, we use $50$ pseudo-observations spread evenly around
the boundary of the zero region, to enforce continuity.  In all three
approaches, there are just two hyperparameters: the variance $\sigma^2$
and the correlation length $\rho$.

\subsection{Prior distribution for the hyperparameters}
\label{sec:prior}

In our application, $d = 2$, the dimension of
the domain, and we choose $\nu = 1$, the shape parameter of the
\Matern covariance function.  This implies $\alpha = 2$, where $\alpha$
is a parameter of the SPDE used to induce the precision
matrix $Q_W$ in \eqref{eq:hier2}.  In the equations below we will
use these explicit values of $d$, $\nu$, and $\alpha$ to simplify
some expressions.

We use the \code{R-INLA} package for computation; see
\citet{lindgren15}.  This package expects the \Matern hyperparameters
$(\sigma, \rho)$ to be specified in terms of the alternative parameters $(\kappa, \tau)$, where 
\begin{equation}\label{eq:sigrho}
  \sigma^2 = \frac{1}{4\pi \, \kappa^2 \, \tau^2 } , \quad 
  \rho = \frac{\sqrt{8}}{\kappa} .
\end{equation}
Under this representation, $\rho$ is the distance at which the
correlation function has fallen to about $0.13$.  \code{R-INLA}
represents the prior for $(\tau, \kappa)$ as 
\begin{subequations}\label{eq:prior}
\begin{align} 
  \log \kappa & = \kappa_0 + a_1\, \theta_1 + a_2\, \theta_2 \\
  \log \tau & = \tau_0 + b_1\, \theta_1 + b_2\, \theta_2
\end{align}
\end{subequations}
where all terms on the right-hand side bar the $\theta$'s are specified,
and $(\theta_1, \theta_2)$ is a Gaussian vector with
specified expectation
and precision.  Thus we must convert our beliefs about $(\sigma,
\rho)$ into values for $(\tau_0, \kappa_0)$, $(a_1, a_2)$, $(b_1,
b_2)$,
and the expectation and precision of $(\theta_1, \theta_2)$, which are modelled as log-normal.

Solving \eqref{eq:sigrho} in logs,
\begin{equation}\label{eq:prior1}
\begin{split}
  \begin{pmatrix}
    \log \kappa \\ \log \tau 
  \end{pmatrix}
  =
  \half \, \begin{pmatrix}
    \log 8 \\ -\log (4\pi) - \log 8
    \end{pmatrix} + \begin{pmatrix}
      0 & -1 \\ -1 & 1
    \end{pmatrix} \,
    \begin{pmatrix}
      \log \sigma \\ \log \rho
    \end{pmatrix}
\end{split} .
\end{equation}
This expression identifies the terms in \eqref{eq:prior}, with
$\theta_1 = \log \sigma$ and $\theta_2 = \log \rho$.

We treat $\sigma$ and $\rho$ as \textit{ a priori} independent.  Our starting point are
prior expectations ${\E(\sigma) = 1.5 \unit{mm/yr}}$ and ${\E(\rho) = 1000
\unit{km}}$. When implemented in \code{R-INLA}, the distance between any two points is represented by the great circle distance on a unit ball; hence $\rho$ need to be scaled by the Earth radius $6371 \unit{km}$, and becomes  ${\E(\rho) = 1000/6371 \approx 0.16}$. For our prior uncertainty, we set the
prior standard deviations to be twice the prior expectations, i.e.\ 
a coefficient of variation of $2$.  If $\log Z \sim \text{N}(m,
s^2)$, then $\E(Z) = \exp(m + s^2/2)$ and $\CV(Z) = \sqrt{\exp(s^2) -
1}$. Hence
\begin{equation}\label{eq:prior2}
  \E \begin{pmatrix} \theta_1 \\ \theta_2 \end{pmatrix}
  = \begin{pmatrix} \log (1.5) - \log \sqrt{5}  \\ \log (2000/6371) - \log \sqrt{5} \end{pmatrix} , \quad 
  \Var \begin{pmatrix} \theta_1 \\ \theta_2 \end{pmatrix}
  = \begin{pmatrix}  \log 5 & 0 \\ 0 & \log 5 \end{pmatrix} ,
\end{equation}
which together with \eqref{eq:prior1} and \eqref{eq:prior2} completes the specification of the prior distribution for the hyperparameters.

\subsection{Results}

\begin{figure}[htbp]
 \centering
\includegraphics[width=\textwidth]{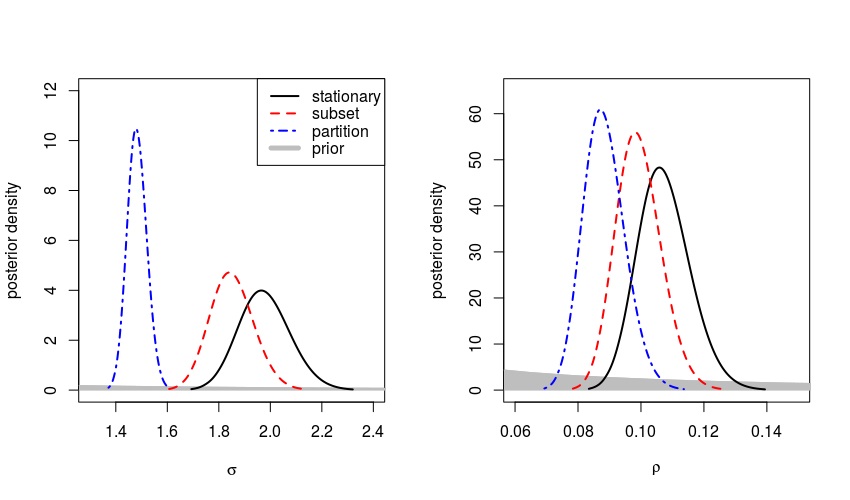}
\caption{Posterior distribution of the hyperparameters. Part of the region below the prior densities are shown as the grey polygons in the plots.}
\label{fig:hyper-par}
\end{figure}

We present the results from the globally stationary model and the two approaches for modelling non-stationarity: the subset model and the constrained parameter partition model.  Figure~\ref{fig:hyper-par} shows the marginal posterior distributions for the two hyperparameters.  Clear differences between these marginal distributions indicate that our different ways of treating non-stationarity are practically as well as theoretically different; although these differences will not necessarily translate into differences in the updated discrepancy.

\begin{figure}[htbp]
 \centering
\includegraphics[width=\textwidth]{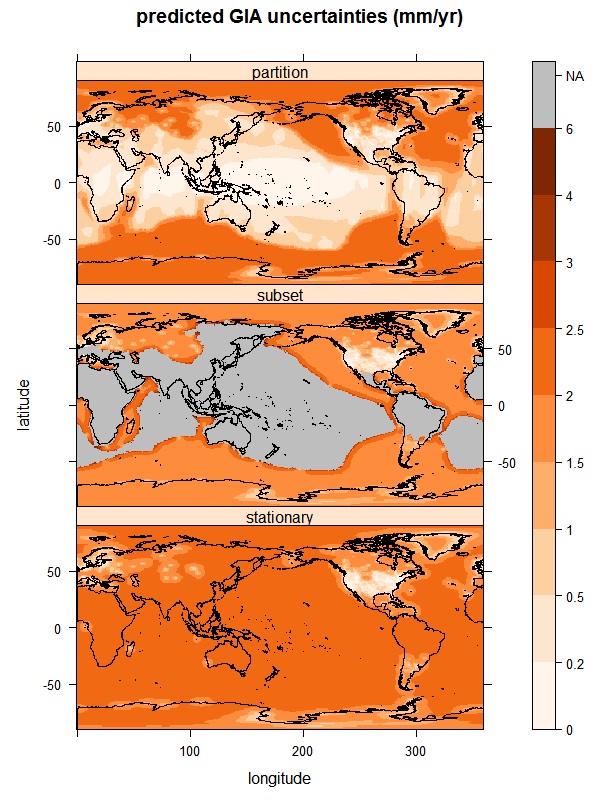}
\caption{Predicted GIA uncertainty. The subset model has no prediction in the zero-region and thus no predicted uncertainties either. The partition model shows much lower uncertainty in the zero-region as expected but the uncertainties are pushed in to the region with less information and thus resulting higher uncertainties in the region of interest compared to the other models. }
\label{fig:uncertainty}
 \end{figure}

Nevertheless, some clear differences are seen in the posterior expectation and standard deviation, shown in Figures~\ref{fig:hyper-par} and \ref{fig:uncertainty}.  In the expectation, the two non-stationary models have higher resolution features relative to the stationary model, e.g., in North America where there are lots of GPS stations: this might reflect the shorter correlation lengths shown in the right panel of Figure~\ref{fig:hyper-par}.  As might be expected, there are larger differences in the standard deviations.  Comparing the two non-stationary models, the presence of an explicit zero-region in the parameter partition model is marked by close-to-zero standard deviations inside the zero-region, and larger standard deviations in the region of interest.  The subset model has an interesting 'Gibbs effect' of a ridge of slightly raised standard deviations just outside the boundary of the zero region, which is completely absent from the parameter partition model.

\begin{figure}[htbp]
 \centering
\includegraphics[width=\textwidth]{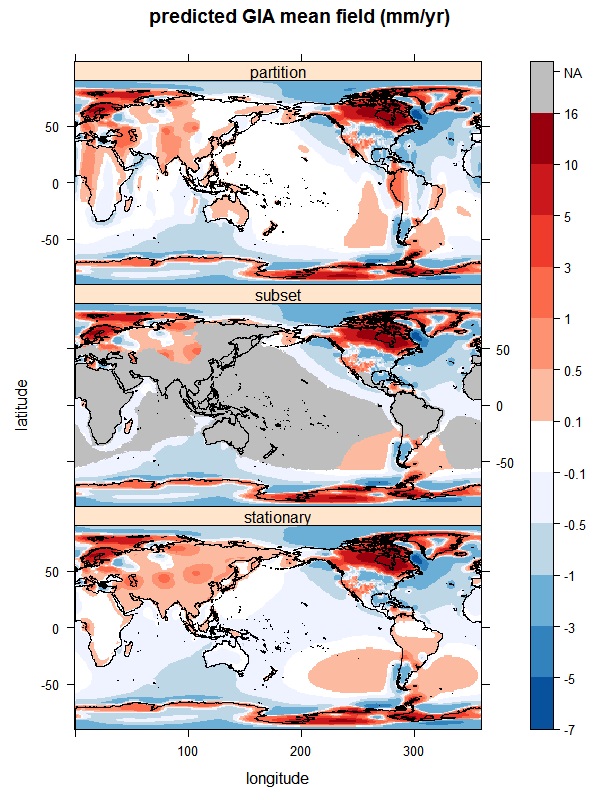}
\caption{Predicted GIA mean field. From top to bottom are the predicted GIA mean field from the constrained partition model, the subset model and the stationary model. The subset model has no prediction in the zero-region. All models performs similarly in the region of interest but show difference in the zero-region and near the boundary.}
\label{fig:mean}
 \end{figure}
 
Looking at both the expectation and the standard deviation, we are much happier presenting the results of the parameter partition model to geoscientists, than the other two models.

\section{Conclusion}
\label{sec:conc}

In this paper we have proposed a Bayesian hierarchical model to synthesize model output and imperfect observations, over a spatial domain in which the discrepancy between the model and the true process has a systematic component.  We model the discrepancy with an isotropic Gaussian random field.

We have addressed two important challenges.  First, the challenge of large-scale computation, which is increasingly common in environmental statistics, where many interesting questions concern global behaviour, and many interesting datasets are dense and global in their coverage.  We have provided a review of the key issues, and the benefits of the SPDE approach proposed by \citet{lindgren11}.  This approach uses a bespoke spatial triangulation, which can be adapted to the dataset and the needs of the inference.

Second, we have proposed a variety of methods for modelling non-stationarity.  Non-stationarity will often be a feature in practice, especially over very large spatial domains which encompass several different types of region.  In this paper we have used what we term a `parameter partition' model.  We have used this model to impose a zero-region on our update of glacio-isostatic adjustment (GIA), which also involves the introduction of pseudo-observations.  In forthcoming work, where it is important to separate land and ocean effects, we will use what we term a `process partition' model.

We have also provided practical guidance, including how to construct triangulations over regions, and how to parameterize the prior distribution of the hyperparameters in the R-INLA package.

The next step for us is the use of multiple latent processes, with more complex observation operators. Some of these observations have large spatial footprints (e.g., gravitation measurements from the GRACE satellite), and, at their native resolution, non-zero measurement error covariances.  This combination of multiple processes and large footprints will push our current computing resources to the limit, and likely require some further approximations.

\section*{Acknowledgments}
The authors are grateful for the fnancial support provided by the European Research Council (ERC) under the European Union's Horizon 2020 research and innovation programme under grant agreement No 69418. We would like to thank H. Bakka and H. Rue for providing technical support on using the R-INLA package.

\bibliography{references}

\end{document}